\documentclass[twocolumn,nonbibnotes,nofootinbib,notitlepage,longbibliography]{revtex4-2}
\usepackage{amsmath}
\usepackage{amssymb}
\usepackage{bm}
\usepackage{epsfig}
\usepackage{graphicx}
\usepackage{xcolor}
\usepackage[colorlinks=true,citecolor=blue,urlcolor=black]{hyperref}
\usepackage[english]{babel}

\usepackage[bbgreekl]{mathbbol}

\def\dfrac{\displaystyle\frac}

\newcommand{\rev}[1]{\textcolor{black}{#1}}

\usepackage{physics}

\newcommand{\h}{\hbar}
\newcommand{\w}{\omega}

\begin{document}

\title{Flexural deformations and collapse of bilayer two-dimensional crystals by interlayer exciton}

\author{Z.A. Iakovlev}
\author{M.A. Semina}
\author{M.M. Glazov}
\affiliation{Ioffe Institute,  	194021 St.~Petersburg, Russia}

\author{E.Ya. Sherman}
\affiliation{Department of Physical Chemistry, 
University of the Basque Country UPV/EHU, 48080 Bilbao, Spain}
\affiliation{IKERBASQUE Basque Foundation for Science, Bilbao, Spain}
\affiliation{EHU Quantum Center, University of the Basque Country UPV/EHU}

\begin{abstract}
We develop a consistent theory of the interlayer exciton-polaron formed 
in atomically thin bilayers. Coulomb attraction between an electron and a hole situated in the 
different layers results in their flexural deformation and provides an efficient mechanism 
of the exciton coupling with flexural phonons. We study the effect of layers tension 
on the polaron binding energy and effective mass leading to suppression of polaron 
formation by the tension both in the weak and strong coupling regimes. We also consider 
the role of the nonlinearity related to the interaction between the out- and in-plane lattice 
displacements and  obtain the criterion of the layer sticking, where the exciton collapses, 
due to the Coulomb attraction between the charge carriers.
\end{abstract}

%\date{\today}

\maketitle

\section{Introduction}

Polarons, compound condensed matter quasiparticles, formed by charge carriers interacting with 
internal excitations such as phonons or magnons, can play a critical role in the understanding 
of the properties of solids ~\cite{Alexandrov2010,emin2013polarons}. 
The polarons formed by electrons and phonons can be responsible 
for conductivity of semiconductors \cite{pekar1946local,landau:pekar,HOLSTEIN1959325,PhysRevLett.74.5144,Alexandrov2010},
their optical properties \cite{Klingshirn2012,vasmono}, and collective electron phases \cite{Krivoglaz1974,PhysRevB.52.1512}. 
Similar effects appear in magnetic semiconductors 
due to interaction between electrons and magnons \cite{PhysRevLett.71.1067}, as well as a result of the hyperfine interaction between the electron and nuclear spins~\cite{Merkulov:1998aa}.
In high-temperature superconductors polarons built by electrons and magnons 
\cite{PhysRevB.37.3759,PhysRevB.58.6194} may determine the magnetic 
properties and the type of superconductivity. Several non-trivial quantum effects in polaron physics were addressed in Refs.~\cite{1991JETPL..53..479G, 1991JETPL..54..285G,Kuklov:1989wh}.

The physics of the polarons depends on three main parameters. First one is 
the coupling strength between the particle and the system excitations, such as phonons or magnons. The second parameter is the spectrum of the excitations dependent on the effective system stiffness. When the host material becomes soft,
the coupling naturally increases. The third parameter is the dimensionality
since in the systems with low dimensionality even relatively weak coupling 
can have a strong effect on the motion
of the particle and ultimately cause its localization. 

The novel atomically-thin semiconductors such as graphene and transition-metal dichalcogenide monolayers,
provided a playground for studies of their rich physical properties, including various aspects of coupling of
the charge carriers to the host layers. In these two-dimensional materials, the coupling of 
phonons to electrons, excitons, and polaritons \cite{PhysRevB.104.L241301} 
provide efficient relaxation channels~\cite{PhysRevB.85.115317,Song:2013uq,PhysRevB.90.045422}, form sidebands 
in absorption and emission spectra~
\cite{PhysRevLett.119.187402,shree2018exciton}, control resonant Raman processes, 
and impact coherence generation~\cite{PhysRevLett.115.117401,PhysRevResearch.1.032007,He:2020aa,PSSR:PSSR201510291}.

Furthermore, one of the key aspects of two-dimensional materials physics is given by out-of-plane 
displacements of atoms forming soft flexural phonons with the parabolic in the wavevector 
dispersion~\cite{ll7_eng,nelson:2004aa,Falkovskii:2012aa,Katsnelson:2013aa} at a large phonon wavelength. 
Due to the low dimensionality and softness, flexural phonons result in rippling and crumpling of the atomic planes, produce anomalous temperature-dependent elasticity ~\cite{PhysRevB.92.155428,Gornyi:2016aa,2019arXiv190702010A,PhysRevE.101.033005}, 
and are of interest for mechanical and optomechanical applications~\cite{Morell:2016aa,Morell:2019aa,Steeneken2021}.
However, it is difficult to achieve a strong coupling of electrons moving in the atomic planes to the out-of-plane displacement flexural phonons  since the interaction of carriers with the flexural phonons 
is suppressed~\cite{M.IKatsnelson01282008,Katsnelson:2013aa,PhysRevB.82.205433} 
as it requires deformations of the single layer along the orthogonal axis, similar to the spontaneous symmetry 
breaking.

\begin{figure}[ht]
    \centering
    \includegraphics[width=\linewidth]{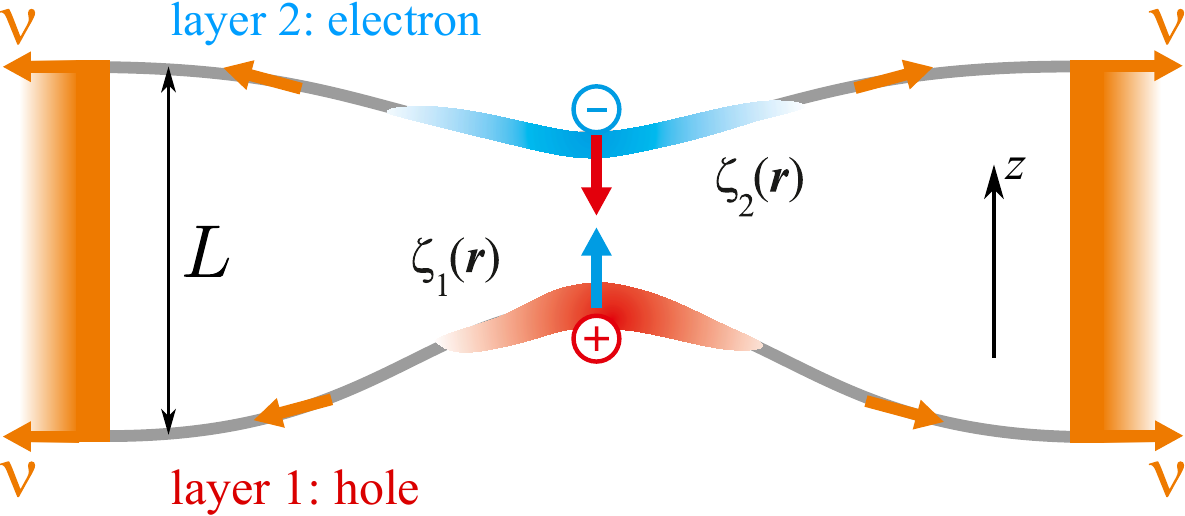}
    \caption{Schematic illustration of the system under study. }
    \label{fig:scheme}
\end{figure}

Recently, bilayer structures attracted a lot of attention of researchers. One of the interesting objects there is an interlayer exciton, where electron and hole, bound by the modified Coulomb interaction, are located in 
different layers. This out-of-plane interaction activates coupling of interlayer excitons to
the flexural phonons since the attraction between electron and hole tends to displace the carriers out of the atomic planes.
Thus, the interlayer excitons, as it was theoretically demonstrated in Ref. \cite{https://doi.org/10.1002/andp.202000339}, interact much stronger with  out-of-plane phonons 
in double-layer structures shown in Fig.~\ref{fig:scheme} than intralayer complexes or single charge carriers. Here the change of the interlayer distance modifies the Coulomb energy of 
electron-hole interaction. This coupling, be it weak or strong, produces the interlayer exciton-polaron.
%An interesting point here is that the system acquires a mixed dimensionality since the two-dimensional motion
%of exciton and lattice displacements appears due to the coupling in the direction perpendicular to the planes. 
Current experimental techniques allow one to manipulate  the spectrum of flexural phonons by applying external tension, 
thus modifying the exciton-phonon coupling and, in turn, properties of the exciton-polarons.

Another important aspect of the problem of exciton-polaron is the nonlinearity. Since the flexural phonons are soft,
the formation of the exciton-polaron can involve many phonon states with a large out-of-plane atomic displacements,
and, thus, go beyond the quadratic in the lattice deformation 
description of the contribution of the lattice to the polaron energy. This nonlinearity can strongly modify 
the properties of the polaron and make them dependent on the size of 
the nano or micro-flake where the exciton is located. 

The interlayer exciton-polaron has been studied theoretically in Ref.~\cite{https://doi.org/10.1002/andp.202000339} 
in the simplified model that neglected tension of the layers, anharmonicity effects, 
and possibility of layer sticking (exciton-polaron collapse) due to the electron-hole attraction. 
In this work we present the consistent theory of the interlayer exciton-polaron 
both in the linear and anharmonic regimes. We show that tension of the layers in the linear regime 
effectively decreases the coupling to the flexural phonon and study the polaron produced by coupling of interlayer exciton to 
the flexural phonons in the presence of a strong tension. In the nonlinear regime we present the equations
describing the deformation of the layers based on the nonlinear theory of elasticity and find the corresponding deformation
caused by the presence of the interlayer exciton. 

The paper is organized as follows. In Sec.~\ref{sec:linear} we concentrate on the linear regime and present the results 
for the interlayer exciton-polaron for the tension-influenced bilayer. We compare exact results with the asymptotics
for the weak and strong tension. In Sec.~\ref{sec:anh} we derive equations for the exciton-polaron energy and layer deformation in a 
strongly nonlinear regime and study their dependence on the system size. We also study system parameters, which controls the collapse of the bilayers.
In Sec.~\ref{sec:discuss} we discuss possible manifestations of the results obtained in
various domains of parameters for materials such as transition metal dichalcogenides and their nanofabricated 
structures. Then, in Sec.~\ref{sec:concl} we present conclusions of our work. Analysis of the effects of the renormalized temperature-dependent phonon dispersion on formation of polarons and calculation of the shape of the atomic layers are presented in the Appendix.

\section{Linear regime}\label{sec:linear}

Following Ref.~\cite{https://doi.org/10.1002/andp.202000339} we consider the two-layer system as shown in Fig.~\ref{fig:scheme} with the charge carriers of opposite signs located in the different layers. 
\rev{The equilibrium distance between the layers $L$ can vary depending on the realization of the system 
from $\lesssim 1$~nm for homo or heterobilayers where the layers are in ``contact'' 
with each other to $10\ldots 100$~nm for drum-like or resonator-like structures where layers 
can be detached, cf. Ref.~\cite{Morell:2016aa,Morell:2019aa,Steeneken2021,Nathamgari:2019vz}}. 
For simplicity, we assume symmetric system and the equation for the ''breathing'' 
mode where the layers move synchronously towards or against each other reads
\begin{equation}
\label{general}
    \frac{\rho}{2}\pdv[2]{\zeta(\bm{r})}{t} + \frac{B}{2}\Delta^2\zeta(\bm{r}) - \frac{T}{2}\Delta\zeta(\bm{r}) + \frac{\rho\w_0^2}{2}\zeta(\bm{r}) = f(\bm{r}).
\end{equation}
Here $\rho$ is the mass density of the layer, $\zeta(\bm r) =\zeta_1(\bm r) -\zeta_2(\bm r)$ is the relative coordinate of the layers, $\bm r$ is the in-plane coordinate, 
$B$ is the bending rigidity of the layer, $T$ is the external tension and $\omega_0$ is the cut-off frequency. The 
latter is related either to the finite size of the \rev{flake-layers, in which case $\omega_{0}$ is the frequency 
of the lowest, fundamental, out-of-plane vibration mode, or the van-der-Waals 
coupling, see Ref.~\cite{https://doi.org/10.1002/andp.202000339} for details}. 
In the right hand side of Eq.~\eqref{general} $f(\bm r)$ is the pressure caused by the 
Coulomb attraction of the electron and hole. Strictly speaking,  the van-der-Waals 
interaction of the layers results in the additional contribution to 
the pressure, $f_{\rm vdW}(\bm r)$. It results in the attraction of the layers even in 
absence of an exciton which is balanced by the arising tension. It  gives rise to 
an equilibrium deformation $\zeta_{\rm vdW}(\bm r)$. In the linear regime, we 
consider the displacements of layers in respect to this equilibrium deformation 
$\zeta(\bm r) \to \zeta(\bm r) - \zeta_{\rm vdW}(\bm r)$. In the geometry 
depicted in Fig.~\ref{fig:scheme} $\zeta_1(\bm r)>0$, $\zeta_2(\bm r)<0$, and $\zeta(\bm r)>0$ 
in the situation where the layers shift towards to each other.

In the absence of excitons Eq.~\eqref{general} admits the plane wave as a solution $\zeta(\bm r) = \zeta_{\bm q} \exp(\mathrm i \bm q \bm r - \mathrm i \omega_{\bm q} t) + {\rm c.c.}$, with $\bm q$ being the phonon wavevector, $\omega_{\bm q}$ is its frequency,
\begin{equation}
\label{disper}
    \w_q = \sqrt{\w_0^2 + \nu q^2 +  \varkappa^2 q^4} \approx
    \begin{cases}
    \w_0, & \text{small}-q,\\
    \sqrt{\nu} q, & \text{intermediate}-q,
    \\
    \varkappa q^2, & \text{large}-q.
    \end{cases}
\end{equation}
Here we used the following notations
\begin{equation}\label{varkappa_nu}
    \varkappa = \sqrt{\frac{B}{\rho}}, \qquad \nu = \frac{T}{\rho}.
\end{equation}
Note that for intermediate values of $q$, where, on the one hand, $q\gtrsim \omega_0/\sqrt{\nu}$, and, on the other hand, $q \lesssim \sqrt{\nu} / \varkappa$, the dispersion of the phonons is governed by the tension, $\omega \propto q$. For smaller wavevectors the phonon frequency cuts-off at $\omega=\omega_0$, and for larger wavevectors the dispersion is quadratic as expected for flexural vibrations.

The exciton envelope wavefunction can be presented in the following form:
\begin{equation}
    \label{exciton:wf}
    \Psi(\bm r, \bm \varrho) = \psi(\bm r) \varphi(\bm \varrho),
\end{equation}
where $\bm r$ is the in-plane position vector of the exciton center of mass, $\bm \varrho$ is the in-plane relative coordinate of the electron-hole motion. 
The function $\varphi(\bm \varrho)$ is governed by electron-hole attraction and satisfies equation
\begin{equation}
\label{exciton:rel}
    -\frac{\hbar^2}{2\mu} \Delta \varphi(\bm \varrho) + V(\bm \varrho,z)\varphi(\bm \varrho)= -E_B \varphi(\bm \varrho).
\end{equation}
Here $\mu=m_e m_h/m$ is the reduced mass of the electron-hole pair with $m=m_e+m_h$ being the exciton translational motion mass and $m_e, m_h$ are the effective masses of the electrons and holes, $E_B$ is the exciton binding energy, $V(\bm \varrho,z)$ is the potential energy of the electron-hole attraction where $z$ is the local distance between the layers, $z=L-\zeta(\bm r)$, with $L$ being equilibrium interlayer distance. 

The exciton center of mass wave function $\psi(\bm r)$ is determined by the polaron effect, i.e., by the deformation of the layers on large scales, which exceed by far the exciton Bohr radius, and by corresponding variation of the exciton energy. We recall that the electron and hole are localized in the respective layers (corresponding envelopes are omitted in Eq.~\eqref{exciton:wf}); thus, the interlayer exciton induced pressure $f(\bm r)$ {in case of sufficiently small displacement, $\zeta(\bm r) \ll L$,} can be recast in the following form
\begin{equation}
    \label{exciton:pressure}
    f(\bm r) = - \int d^2 \varrho |\varphi(\bm \varrho)|^2 \left.\frac{\partial V(\bm \varrho, z)}{\partial z}\right|_{z=L} |\psi(\bm r)|^2.
\end{equation}
Physical meaning of Eq.~\eqref{exciton:pressure} is clear: $f(\bm r)$ is the normal component of the Coulomb pressure
attracting the electron and hole, related to the $-\partial V/\partial z$ derivative, 
averaged over the exciton envelope function. 
Making use of Eq.~\eqref{exciton:pressure} we obtain the following 
exciton-phonon interaction Hamiltonian
\begin{equation}
   \hat{U} = \sum_{\bm q} U_{\bm q} e^{\mathrm i \bm q {\bm r}} \hat{b}_{\bm q} + {\rm h.c.}, \quad U_{\bm q} = D{\mathcal F}_{s}(q)\sqrt{\frac{\hbar}{\rho \omega_q \mathcal S}}.
\end{equation}
Here $\hat b_{\bm q}$, $\hat b^\dag_{\bm q}$ are the  annihilation and creation operators of the phonons with the wavevector $\bm q$, and the interaction parameters $U_{\bm q}$  are expressed in the standard way via the elementary displacement, $\sqrt{\hbar/\rho \omega_q\mathcal S}$ 
\rev{with $\mathcal S$ being the normalization area}, the ``deformation potential parameter''
\begin{equation}
\label{D:def}
    D = \int d^2 \varrho |\varphi(\bm \varrho)|^2 \left.\frac{\partial V(\bm \varrho, z)}{\partial z}\right|_{z=L}, 
\end{equation}
and the form-factor accounting for the finite Bohr radius of the exciton
\begin{multline}
    \mathcal F_s(\bm{q}) = \frac1{D}\int d^2 \varrho |\varphi(\bm \varrho)|^2\left.\frac{\partial V(\bm \varrho, z)}{\partial z}\right|_{z=L}  \\ \times\frac1{2}\left[\exp(-\mathrm i\bm{q\varrho}\frac{m_h}{m_e+m_h}) + \exp(\mathrm i\bm{q\varrho}\frac{m_e}{m_e+m_h})\right].
\end{multline}

\begin{figure}[t]
    \centering
    \includegraphics[width=\linewidth]{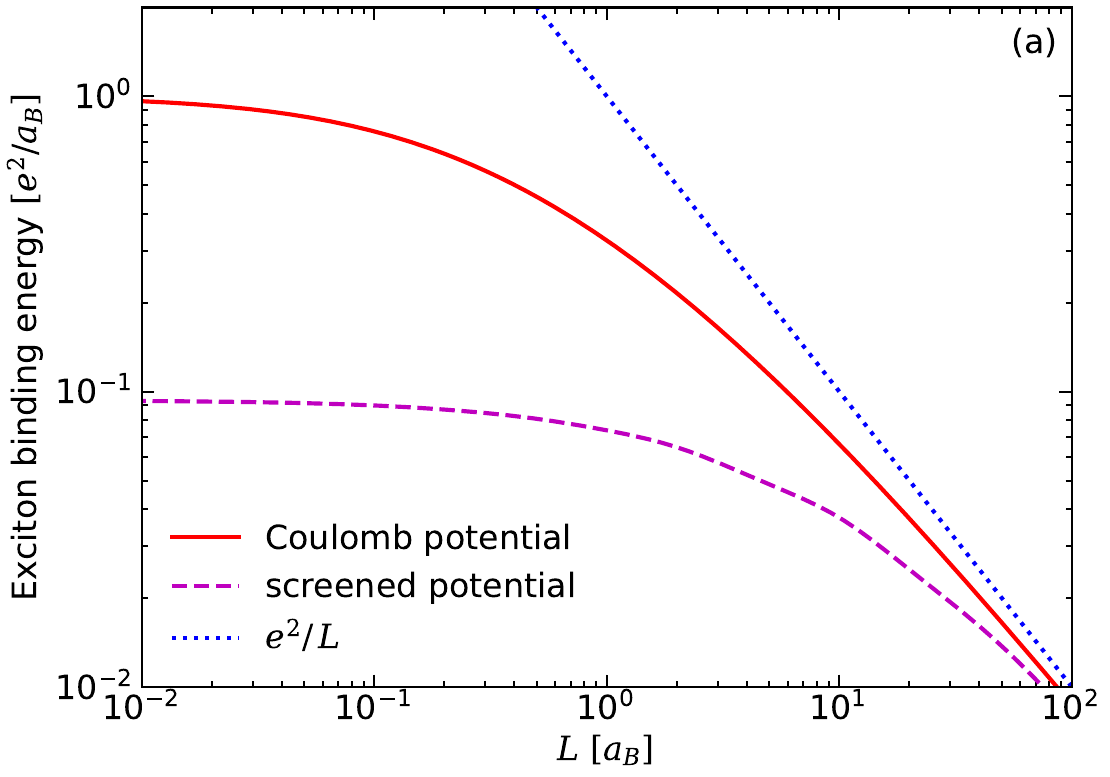}\\
    \includegraphics[width=\linewidth]{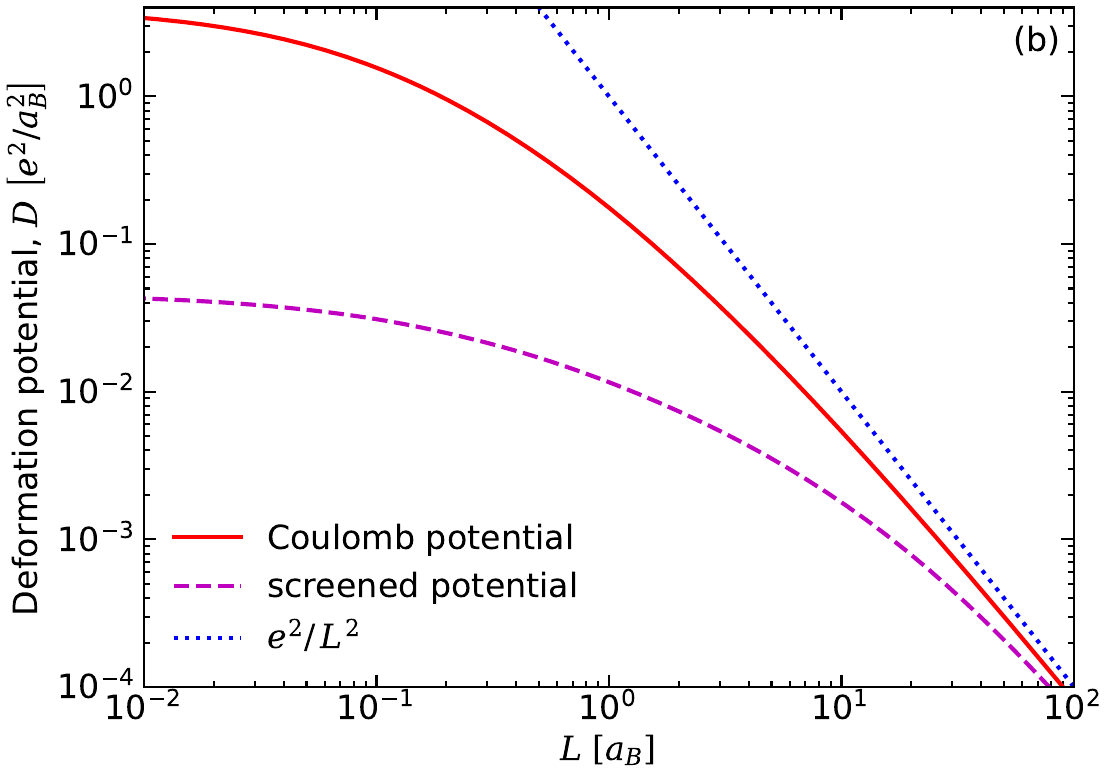}\\
    \caption{Interlayer exciton binding energy $E_B$ calculated after Eq.~\eqref{exciton:rel} [panel (a)] and deformation potential parameter $D$ [panel (b)] calculated after Eq.~\eqref{D:def} as a function of interlayer distance. Solid red curves correspond to the unscreened Coulomb interaction, dashed magenta curves correspond to the screened potential after Ref.~\cite{Semina:2019aa} with the screening radii being $r_{1}=r_{2}=6a_{B}$, dotted blue curves show large-$L$ asymptotics. An extended range of the $L-$ axis is used to illustrate the asymptotics.}
    \label{fig:EbD}
\end{figure}

For the electron-hole interaction in the unscreened Coulomb form, $V({\bm \varrho},z)=-e^2/\sqrt{\varrho^2+z^2}$, the deformation potential parameter can be written, in the limits of small and large interlayer distances as compared to the two-dimensional exciton Bohr radius $a_B=\hbar^2/(2\mu e^2)$, as~\cite{https://doi.org/10.1002/andp.202000339}
\begin{equation}
\label{D:est}
D= \begin{cases}
        {4e^{2}}/{ a_{B}^{2}}, \quad L \ll a_B,\\
        \\
        {e^{2}}/{ L^{2}}, \quad L \gg a_B.
    \end{cases}
\end{equation}
Figure \ref{fig:EbD} shows the dependence of the exciton binding energy $E_B$ [panel (a)] and the deformation potential parameter $D$ [panel (b)] on the interlayer distance $L$ calculated numerically following Ref.~\cite{Semina:2019aa}. Solid red curves show the results for the Coulomb form of the interaction, dashed magenta curves show the results for the potential taking into account dielectric screening in the system (being the extension of the Rytova-Keldysh potential for bilayer, Ref.~\cite{Semina:2019aa}) and dotted blue curves show universal large distance asymptotics, $e^2/L$ and $e^2/L^2$, for the binding energy and deformation potential, respectively.

\subsection{Weak coupling}

In a weak coupling regime the polaron energy is described by the second order perturbation theory. For negligibly small temperatures, there are no phonons in the system, and we need to take into account only virtual phonon emission and absorption process with the result:
\begin{equation}
    \label{2nd:total}
    \delta E = -\sum_{\bm q} \frac{|U_q|^2}{\h\w_q+\h^2(q^2-2\bm{kq})/2m}.
\end{equation}
Evaluating sum over $\bm q$ in Eq.~\eqref{2nd:total} at $\bm k=0$ we  obtain the closed-form expression for the corresponding polaron energy:
\begin{multline}
\label{pol:energ:weak}
    \delta E_\text{w} = -\frac{\beta\h\omega_0}{2\pi}\frac1{\sqrt{1-\varkappa^2/\mathcal{K}^2 + \alpha^2}}\\
    \times
    \tanh^{-1}{\frac{\sqrt{1-\varkappa^2/\mathcal{K}^2 + \alpha^2}}{1+\varkappa/\mathcal{K} + \alpha}}.
\end{multline}
Here we use the following notations:
\begin{equation}\label{beta_alpha_K}
    \beta = \frac{2mD^2}{\rho\h^2\w_0^2}, \quad \alpha = \frac{\nu}{2\w_0\mathcal{K}}, \quad \mathcal{K} = \frac{\h}{2m},
\end{equation}
with $\beta$ being the dimensionless coupling constant, $\alpha$ being the dimensionless tension, and $\mathcal K$ characterizes the steepness of the exciton dispersion. In a weak coupling regime studied in this section the condition $\beta \ll 1$ fulfilled. 

\begin{figure}[ht]
    \centering
    \includegraphics[width=\linewidth]{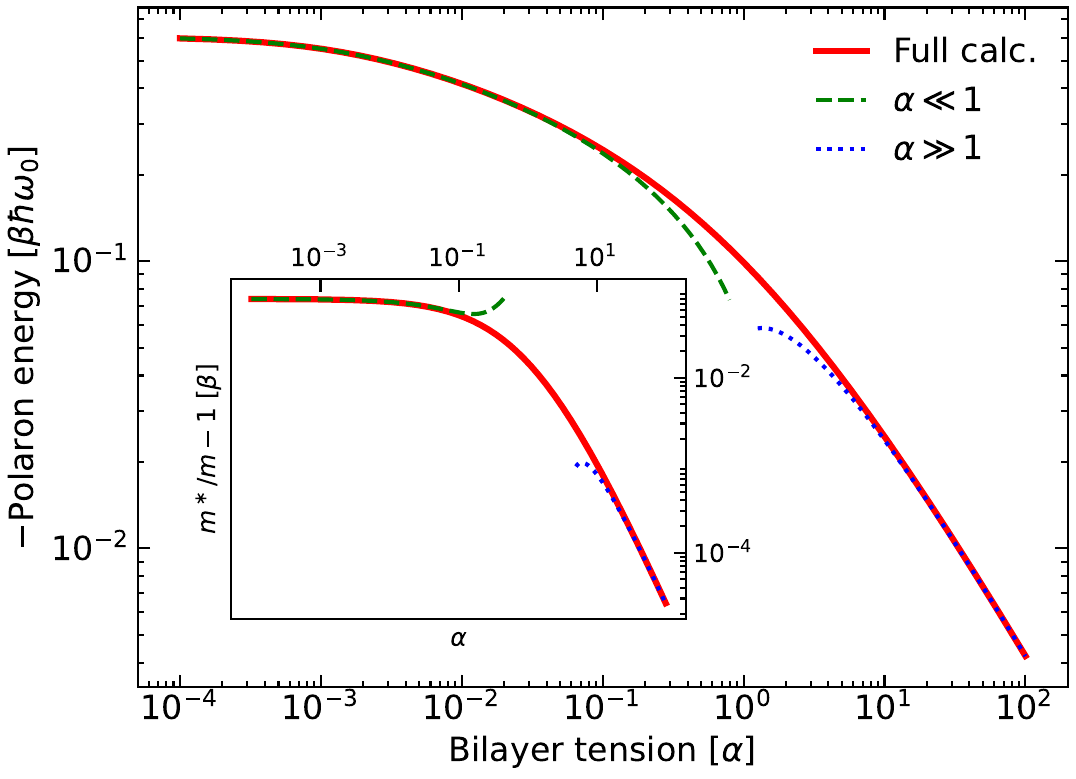}
    \caption{Absolute value of the polaron energy in the units of $\beta\hbar\omega_0$ as function of the bilayer tension. Solid red curve has been calculated after Eq.~\eqref{pol:energ:weak}, dashed green and dotted blue curves show, respectively, small and large tension asymptotics calculated after Eq.~\eqref{asympt}. Here the dimensionless bending rigidity $\varkappa / \mathcal K = 0.001$.  Inset demonstrates the renormalization of the polaron effective mass, $m^{*}/m-1$, in the units of $\beta$ as function of the bilayer tension. Solid red curve shows the full dependence, Eq.~\eqref{pol:mass:weak}, dashed green and dotted blue curves demonstrate the asymptotics, Eq.~\eqref{asympt:m}. An extended range of the $\alpha-$ axis is used to illustrate the asymptotics.}
    \label{fig:weak_coupling}
\end{figure}

The polaron energy in a weak coupling regime as a function of the tension $\alpha$ is shown in Fig.~\ref{fig:weak_coupling} together with its asymptotic values at the small and large tension. They can be readily evaluated from Eq.~\eqref{pol:energ:weak}  as 
\begin{equation}
    \label{asympt}
    \delta E_\text{w} \approx  -\frac{\beta\h\w_0}{4\pi}
    \begin{cases}
        -\ln\left(\varkappa/\mathcal{K} + \alpha\right),  \qquad \alpha \ll 1,\\
        \\
     \alpha^{-1}\ln{2\alpha},  \qquad \alpha \gg 1.
    \end{cases}
\end{equation}

In order to find the exciton-polaron effective mass we decompose the general expression~\eqref{2nd:total} up to the $k^2$ terms and, after some transformations, arrive at
\begin{equation}
\label{pol:mass:weak}
    \frac{m^*}{m} = 1 + \frac{\beta}{4\pi}\xi(\alpha),
\end{equation}
with the function $\xi(\alpha)$ given by
\begin{equation}
   \xi(\alpha) = \frac{1+3\alpha-2\alpha^2}{(\alpha^2+1)^2} - 2\frac{2\alpha-\alpha^3}{(\alpha^2+1)^{5/2}}\tanh^{-1}{\frac{\sqrt{\alpha^2+1}}{1+\alpha}}.
\end{equation}

Correspondingly, in the small- and large-$\alpha$ regimes we have
\begin{equation}
\label{asympt:m}
    \xi(\alpha) \approx \left\{\begin{array}{cc}
        1 - 2\alpha\ln(2/\alpha) + 3\alpha, & \qquad \alpha \ll 1,\\
        \\
        (\ln{2\alpha} - 2) / \alpha^2, & \qquad \alpha \gg 1.
    \end{array}\right.
\end{equation}
The correction to the free exciton effective mass $m^{*}/m-1$ in the units of the coupling constant $\beta$ is plotted in the inset in Fig.~\ref{fig:weak_coupling}.

In the limit of a very small tension, $\alpha\to 0$, our results correspond to 
those obtained in Ref.~\cite{https://doi.org/10.1002/andp.202000339}. Both energy of 
the polaron and renormalization of mass are proportional to the small 
coupling constant $\beta \ll 1$. Polaron energy is logarithmically large due 
to the parameter $\mathcal{K}/\varkappa \gg 1$.  With the increase of the 
tension, the polaron binding energy and correction to the effective mass decrease. 
It is because the system becomes more rigid and the Coulomb attraction between the 
electron and the hole results in smaller deformations of the layers. If the tension 
becomes large, $\alpha \gg 1$, the properties of the system are controlled solely 
by the tension resulting in relation $\omega_q\propto q$. Therefore, the polaron 
energy and mass are independent of the layer bending rigidity $\varkappa$ in this case.

\subsection{Strong coupling}

If the cut-off frequency $\omega_0$ becomes sufficiently small, the dimensionless 
coupling constant $\beta$ becomes larger than unity and the perturbation theory approach presented above becomes inapplicable. 
\rev{It can happen for sufficiently large interlayer separation $L$
in a drum-like structures, see estimates in Ref.~\cite{https://doi.org/10.1002/andp.202000339} 
or, if the cut-off is provided by the finite size of the flake, for sufficiently large flakes.}  

Here we address the strong coupling regime where $\beta \gg 1$ and the vibrations of the 
layers can be described classically. In this regime, the exciton generates 
considerable deformation of the layers and gets self-trapped in the induced potential well. 
Thus, the polaron energy can be found from the minimization of the $\delta E_s$ 
defined as~\cite{https://doi.org/10.1002/andp.202000339}
\begin{equation}
    \delta E_s = -\frac{\h^2}{2m}\left\langle \psi(\bm r) \left| \Delta\right| \psi(\bm r)\right\rangle - \sum_{\bm q} \frac{|U_q|^2}{\hbar\omega_q} \left|F(\bm q)\right|^2,
    \label{strong:total}
\end{equation}
where $\psi(\bm{r})$ is exciton center-of-mass wavefunction and 
\begin{equation}
    F(\bm q) = \left\langle \psi(\bm r) \left| e^{-\mathrm i \bm q\bm r} \right| \psi(\bm r)\right\rangle,
\end{equation}
is the Fourier component of the corresponding probability density. The first term in Eq.~\eqref{strong:total} describes the kinetic energy of the localized exciton and the second term presents the mechanical energy of the layers. The latter is presented as a sum of energies of individual oscillators with the momentum $\bm q$ and frequency $\omega_q$.

Variational approach is used to minimize energy in Eq.~\eqref{strong:total}. A Gaussian trial function 
\begin{equation}
\label{trial:gauss}
    \psi(r) = \sqrt{\frac{2b}{\pi}}e^{-br^2},
\end{equation}
with the polaron size $R = b^{-1/2}$ is assumed. Combining Eqs.~\eqref{strong:total} and \eqref{trial:gauss} we obtain the following expression for the polaron energy
\begin{equation}
    \delta E_s(b) = \frac{\hbar^2 b}{m} - \frac{D^2}{4\pi\rho}\int_{0}^\infty 
    \frac{\exp[-x/(4b)]}{\w_{0}^{2} + \nu x + \varkappa^2 x^2}dx.
    \label{strong:exact}
\end{equation}

In order to obtain an analytical approximation we replace the exponent in the integral in Eq.~\eqref{strong:exact} by the constant but limit the integration over $x$ by $4b$. As a result, the polaron energy can be recast as:
\begin{equation}
    \delta E_s \approx \frac{\hbar^2 b}{m} - \frac{D^2}{4\pi\rho}\int_0^{4b} \frac{1}{\w_0^2 + \nu x + \varkappa^2 x^2}dx.
    \label{strong:E(b)}
\end{equation}
The integral in Eq.~\eqref{strong:E(b)} can be readily expressed via the inverse trigonometrical functions and the minimization over $b$  can be performed analytically (as presented in Appendix B) with the result:
\begin{equation}
    \delta E_s = -\frac{\mathcal{K}}{\varkappa}\frac{\beta\h\w_0}{4\pi}g(\alpha,\beta),
    \label{strong:energy}
\end{equation}
with function $g(\alpha,\beta)$ being
\begin{multline}
    g(\alpha,\beta) 
    = \frac1{\sqrt u}\left(\arctan\frac{\sqrt{{u}}}{{v}} - \arctan\sqrt\frac{u}{\beta/2\pi -u} \right)\\
    - \frac{2\pi}{\beta}\left(\sqrt{\frac{\beta}{2\pi} - u} - v\right),
    \label{full:g(a,b)}
\end{multline}
and $v = \alpha\mathcal K /\varkappa$, $u=1-v^2$.
% \begin{multline}
%     g(\alpha,\beta) = \frac1{\sqrt{1-\left(\frac{\mathcal{K}\alpha}{\varkappa}\right)^2}}\left\{\tan^{-1}\frac{\sqrt{1-\left(\frac{\mathcal{K}\alpha}{\varkappa}\right)^2}}{\frac{\mathcal{K}\alpha}{\varkappa}} \right. \\ \left.- \tan^{-1}\sqrt{\frac{1-\left(\frac{\mathcal{K}\alpha}{\varkappa}\right)^2}{\frac{\beta}{2\pi} - 1 + \left(\frac{\mathcal{K}\alpha}{\varkappa}\right)^2}}\right\}  \\ - \frac{2\pi}{\beta}\left\{\sqrt{\frac{\beta}{2\pi} - 1 + \left(\frac{\mathcal{K}\alpha}{\varkappa}\right)^2} - \frac{\mathcal{K}\alpha}{\varkappa}\right\}.
%     \label{full:g(a,b)}
% \end{multline}

\begin{figure}[ht]
    \centering
    \includegraphics[width=\linewidth]{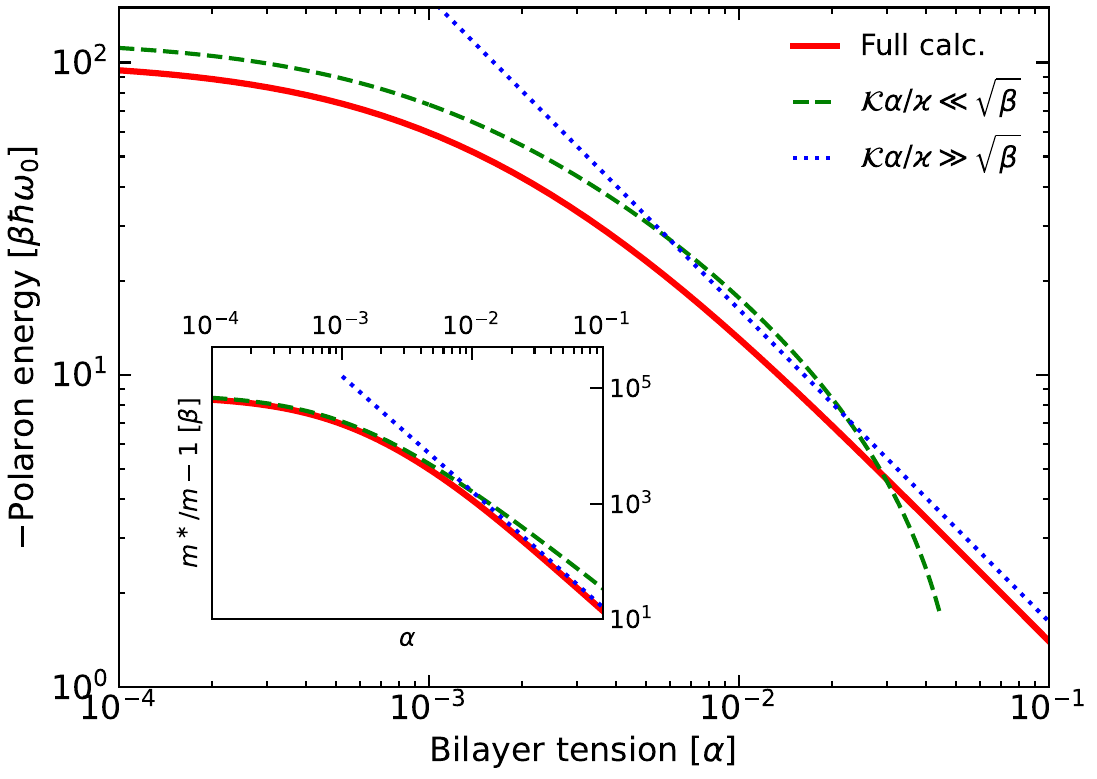}
    \caption{Absolute value of the polaron energy in the strong coupling regime in the units of $\beta\hbar\omega_0$ as function of the bilayer tension. Solid red curve has been calculated after Eq.~\eqref{strong:exact}, dashed green and dotted blue curves show, respectively, small and large tension asymptotics calculated after Eq.~\eqref{strong:asympt}. Here the dimensionless bending rigidity $\varkappa / \mathcal K = 0.001$ and dimensionless coupling parameter $\beta = 1000$. Inset demonstrates the renormalization of the polaron effective mass, $m^*/m-1$, in the units of $\beta$ as function of the bilayer tension. Solid red curve shows the full dependence, Eq.~\eqref{mass_exact}, dashed green and dotted blue curves demonstrate the asymptotics, Eq.~\eqref{asympt:m_strong}.}
    \label{fig:strong_coupling}
\end{figure}

Figure~\ref{fig:strong_coupling} demonstrates the polaron energy in the strong coupling regime as a function of tension $\alpha$ and its asymptotic behavior at  small and large tension. Analytical approximation of energy, Eq.~\eqref{full:g(a,b)}, is very close to the result of exact numerical calculation (within the thickness of the curves in Fig.~\ref{fig:strong_coupling}) and is not shown. The limits of small and large tension can be found from Eq.~\eqref{strong:energy} as
% \begin{equation}
%     g(\alpha, \beta) \approx 
%     \begin{cases}
%         \frac1{\sqrt{u}}\arctan{\frac{\sqrt{u}}{v}} - \sqrt{\frac{2\pi}{\beta}}, & \qquad v \ll \sqrt{\beta},\\
%         \\
%      \frac1{2v}\left(\ln\frac{\beta}{2\pi} - 1\right), & \qquad v \gg \sqrt{\beta}.
%     \end{cases}
%     \label{strong:asympt}
% \end{equation}
 \begin{equation}
    g(\alpha, \beta) \approx 
    \begin{cases}
        \displaystyle{\frac{1}{\sqrt{u}}\arctan{\frac{\sqrt{u}}{v}} - \sqrt{\frac{2\pi}{\beta}}}, & \qquad v \ll \sqrt{\beta},\\
        \\
      \displaystyle{\frac{1}{2v}\left(\ln\frac{\beta}{2\pi} - 1\right)}, & \qquad v \gg \sqrt{\beta}.
    \end{cases}
    \label{strong:asympt}
\end{equation}
% \begin{equation}
%     g(\alpha, \beta) \approx 
%     \begin{cases}
%       u^{-1/2}\arctan{u^{-1/2}/v} - \sqrt{\frac{2\pi}{\beta}}, & \qquad v \ll \sqrt{\beta},\\
%         \\
%      \frac1{2v}\left(\ln\frac{\beta}{2\pi} - 1\right), & \qquad v \gg \sqrt{\beta}.
%     \end{cases}
%     \label{strong:asympt}
% \end{equation}
%  \rev{\begin{equation*}
%      g(\alpha, \beta) \approx 
%      \begin{cases}
%          \arctan\left(\sqrt{u}/v\right)/\sqrt{u} - \sqrt{{2\pi}/{\beta}}, & \quad v \ll \sqrt{\beta},\\
%          \\
%       {\left(\ln\left({\beta}/({2\pi})\right) - 1\right)}/{2v}, & \quad v \gg \sqrt{\beta}.
%      \end{cases}
%      \label{strong:asympt}
%  \end{equation*}}
At $\alpha\to 0$ the expressions Eq.~\eqref{strong:energy} pass to formulas derived in Ref.~\cite{https://doi.org/10.1002/andp.202000339}. Similarly to the weak coupling regime, the tension results in a decrease of the polaron energy. Note that polaron bound state within the strong coupling approach exists only at $\beta \gg 1$, which is, in reality, the condition for the realization of said regime by definition.

Polaron effective mass can be expressed as~\cite{https://doi.org/10.1002/andp.202000339}
\begin{equation}
    \frac{m^{\ast}}{m} - 1 = \sum_{\bm q}\frac{f_{\bm q}^2}{\rho \omega_{q}^4}\frac{q^2}{m} = \frac{\beta}{{4}\pi}\left(\frac{\mathcal K}{\varkappa}\right)^2h(\alpha, \beta),
    \label{mass_exact}
\end{equation}
where $f_{\bm q}$ is  Fourier component of the pressure. After transformations similar to the ones used to obtain Eq.~\eqref{strong:energy} from Eq.~\eqref{strong:exact} we arrive at the analytical expression for renormalization of mass
\begin{multline}
    h(\alpha, \beta) = \frac1{u}\left[1 - \frac{u + v\sqrt{\beta/(2\pi) - u}}{\beta / (2\pi)} \right. \\ \left. - \frac{v}{\sqrt{u}}\left(\arctan\frac{\sqrt{u}}{v} - \arctan\sqrt{\frac{u}{\beta / (2\pi) - u}}\right)\right].
    \label{full:h(a,b)}
\end{multline}

Inset in Fig.~\ref{fig:strong_coupling} shows correction to the polaron mass in strong coupling regime Eq.~\eqref{mass_exact}. An analytical approximation Eq.~\eqref{full:h(a,b)} is in good agreement with the numerical one obtained using the exact expression Eq.~\eqref{mass_exact}. In cases of small and large tension, Eq.~\eqref{full:h(a,b)} can be simplified as
\begin{equation}
    h(\alpha, \beta) \approx \begin{cases}
        \dfrac{1}{u}\left(1 - \dfrac{v}{\sqrt{u}}\arctan{\dfrac{\sqrt{u}}{v}}\right), & \qquad v \ll {\sqrt{\beta}}, \\
        \\
        \dfrac{1}{2v^{2}}\left(\ln{\dfrac{\beta}{2\pi}} - 1\right), & \qquad v \gg {\sqrt{\beta}}.
    \end{cases}
    \label{asympt:m_strong}
\end{equation}
% \rev{\begin{equation*}
%     h(\alpha, \beta) \approx \begin{cases}
%         \left(1 - {v}\arctan\left(\sqrt{u}/v\right)/{\sqrt{u}}\right)/u, & \qquad v \ll \beta, \\
%         \ln\left(\beta/({2\pi}) - 1\right)/(2v^{2}), & \qquad v \gg \beta.
%     \end{cases}
%     \label{asympt:m_strong}
% \end{equation*}}
For sufficiently small tension, $\alpha \ll \varkappa\sqrt{\beta}/ \mathcal K$, the polaron mass is parametrically larger than mass of free exciton, because exciton drags a significant deformed area of the layers.

Interestingly, for the large enough tension, both in the weak coupling and strong coupling regimes ($\alpha \gg 1$ for weak coupling and $\alpha \gg \varkappa {\sqrt{\beta}} / \mathcal K$ for strong coupling regime), energy and mass of polaron do not depend on the bending rigidity $\varkappa$ and are related by
\begin{equation}
m^\ast - m = - \rho \delta E /{T}.
%    \rev{\delta E = -\left(\frac{m^\ast}{m} - 1\right)\alpha\hbar\omega_0 = -(m^\ast - m)\nu = -\frac{m^\ast - m}{\rho}I}.
\end{equation}
In fact, $-\delta E /T$ (note that $\delta E<0$) is the area where the exciton-induced deformation is significant and $-\rho \delta E /{T}$ is its mass.

%\commentZakhar{It looks like the area of deformed part of the layer multiplied by the tension. It seems like $-$ characteristic strain energy of the part of the layer with polaron.}

\section{Anharmonic regime}\label{sec:anh}

\begin{figure}[ht]
    \centering
    \includegraphics[width=\linewidth]{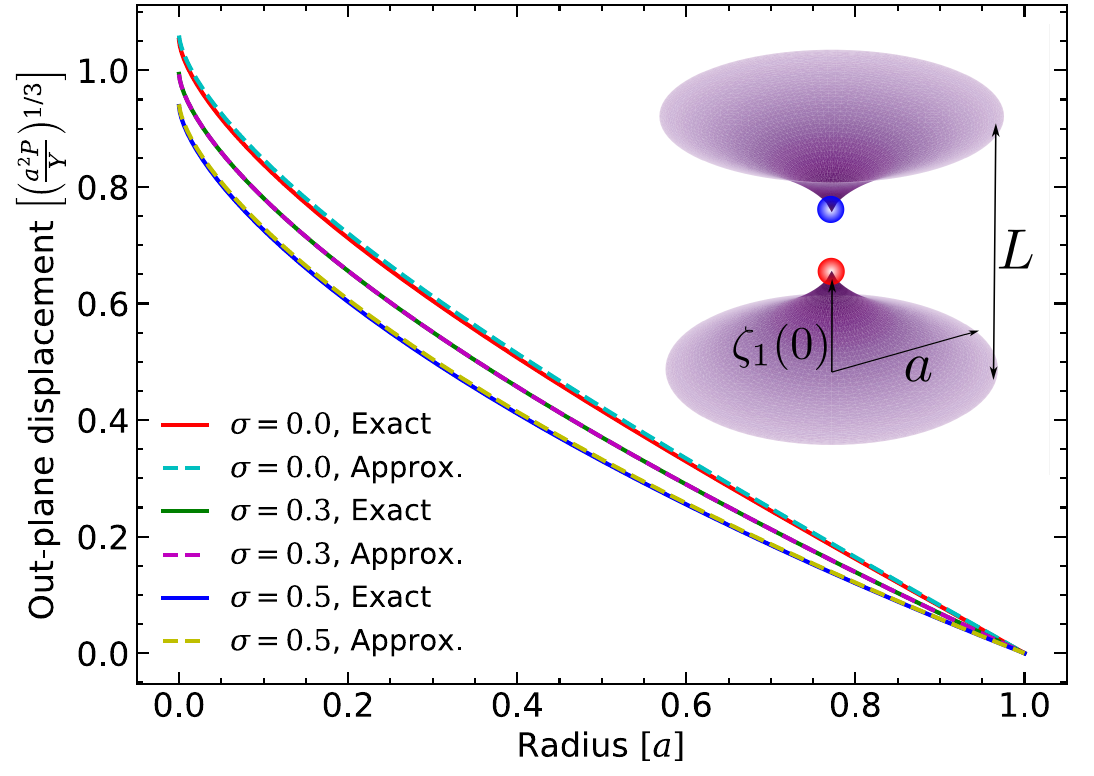}
    \caption{Shape of the layer in anharmonic regime with point force at the center of circular flake for different Poisson's ratio. Solid lines show exact solution of Eqs.~(\ref{system})-(\ref{boundary_cond}). Dashed lines demonstrate approximate form of the layers Eq.~(\ref{form}). Inset illustrates the bilayer exciton-polaron in nonlinear regime.}
    \label{fig:3Dscheme}
\end{figure}

As we demonstrated above, the polaron energy and mass are proportional to the dimensionless coupling constant $\beta$ and, since $\beta \propto \omega_0^{-2}$ diverge if the cut-off frequency $\omega_0$ introduced in Eqs.~\eqref{general} and \eqref{disper} diminishes. The cut-off frequency $\omega_0$ can be related to the in-plane size of the structure or to the van-der-Waals interaction between the layers. For sufficiently large system and interlayer distances the coupling constant $\beta$ can become so large that within presented model the spatial extension of the deformed area becomes comparable with the in-plane size of the structure. In this situation, linear analysis developed in the previous sections becomes invalid and we have to consider  anharmonic contributions. Since said contributions are important  only for sufficiently large $\beta$, the vibrational modes of the layers can be treated classically. In contrast to the previous section, here we go beyond the harmonic approximation for the flexural mode and take into account that, under flexing, an in-plane deformation of the layer occurs. The schematic illustration of the bilayer exciton-polaron is shown in the inset to Fig.~\ref{fig:3Dscheme}. 
%\rev{\sout{Here we use the theory of the anharmonic deformations of the thin membranes
%where free-energy due to the deformations is recast as (we omit the explicit ${\bm r}-$dependence for brevity):}}
%\begin{multline}
%\label{Free:energy}
%\Phi = \int d\bm \varrho \Bigl\{\sum_{l=1,2} \left[B (\Delta \zeta_l)^2 + \lambda u_{l;ii}^2 +2\mu %u_{l;ij}u_{l;ij}\right]\\
%- (\zeta_1-\zeta_2)f\Bigr\},
%\end{multline}
%\rev{\sout{where $B$ is the  bending rigidity (cf. Eq. \eqref{general}), $\lambda$ and $\mu$ are the Lam\'e constants, 
%$u_{l;ij}$ are the in-plane components of the strain tensor of the layer $l=1,2$ and $i,j=x,y$. }}

Following Refs.~\cite{ll7_eng,Katsnelson2020} we obtain the equilibrium conditions in the form 
of the axially-symmetric F\"oppl-von K\'arm\'an equations assuming that the flake has a circular shape:
\begin{eqnarray}
    &&\frac{1}{r}\hat{L}_{r}\frac{d\varphi}{dr} + \frac{Y}{2}\left(\frac{d\zeta_{1}}{dr}\right)^2 = 0,    \label{foppl} \\
    && B\hat{L}_{r}\frac{d\zeta_{1}}{dr} - r\frac{d\varphi}{dr}\frac{d\zeta_{1}}{dr} = \frac{Pr}{2\pi}. \nonumber    
\end{eqnarray}
Here the differential operator $\hat{L}_{r}$ is defined as
\begin{equation}
\label{eq:Lr:init}
    \hat{L}_{r}\chi = \left(r^2\frac{d^2}{dr^2} + r\frac{d}{dr} - 1\right)\chi = r^2\frac{d}{dr}\left(\frac{1}{r}\frac{d}{dr}\left(r\chi\right)\right),
\end{equation}
where $\chi\equiv \chi(r)$ is an arbitrary function of the radius.
Other notations are follows:  $\zeta_{1}\equiv\zeta_{1}(r)$ is an out-plane displacement of the lower layer 
(the displacement of the upper layer has the same absolute value but opposite sign and satisfies the analogous equation), 
$\varphi\equiv\varphi(r)$ is the Airy stress function, $Y$ is the two-dimensional Young's modulus, and
\begin{equation}
\label{force}
\rev{P \equiv P\left(\zeta_1(0)\right)}=\frac{e^2}{[L - 2\zeta_1(0)]^2}
\end{equation}
is a force applied to the layer due to the attraction of the charge carrier in this layer to the carrier in the  other layer. In Eq.~\eqref{force} we assumed that the exciton Bohr radius is much smaller than the interlayer distance $L$ and the size of the deformed area, which makes it possible to neglect the in-plane extension of the exciton and consider it as point-like and use Coulomb form of their interaction. Note that in our coordinate frame (see Fig.~\ref{fig:3Dscheme}) the $z$-axis points from the bottom to the top layer, correspondingly, $\zeta_{1}>0$, and $P>0$.

It is convenient to make the following substitutions
\begin{equation}
    g = \left(\frac{2\pi aY}{P}\right)^{1/3}\frac{d\zeta_{1}}{dr};\quad h = \left(\frac{2\pi aY}{P}\right)^{2/3}\frac1{aY}\frac{d\varphi}{dr},
    \label{fg}
\end{equation}
and introduce the dimensionless coordinate $x = r/a$ (we recall that $a$ is the radius of the system). As a result, we obtain
\begin{equation}
      \hat{L}_{x}h + \frac{x}{2}\frac{1}{h^2} = 0.
      \label{system}
\end{equation}
In derivation of Eqs.~\eqref{foppl} we have also disregarded the bending rigidity, i.e., we assumed that
\begin{equation}
    \left(\frac{4\pi^2B^3}{a^4P^2Y}\right)^{1/3} \to 0,
    \label{anharmonic:limit}
\end{equation}
which also yields $g=-1/h$.
Note that in the opposite case, where $B$ is sufficiently large we recover linear regime studied above. 

Nonlinear Eqs.~\eqref{foppl} and \eqref{system} should be supplemented with the boundary conditions, which we take in the form of absence of the in-plane displacements at the center of the layer and at the edges. Thus, the boundary conditions can be formulated as
\begin{eqnarray}   
&&        xh^{\prime}{(x)} - \sigma h{(x)}  \to  0, \text{ at } x \to 0,  \label{boundary_cond}\\
&&        h^{\prime}(1) - \sigma h(1)  =  0. \nonumber 
\end{eqnarray}
Here $\sigma$ is the  Poisson's ratio. Noteworthy, due to the axial symmetry of the problem, it affects only the boundary conditions. 
%In this situation after passing to the dimensionless coordinates, $x = r/a$, we arrive at
%\begin{equation}
%    \begin{cases}
%        g = 1/f \\
%        L_xf + \frac{x}{2}\frac1{f^2} = 0
%    \end{cases}
%   \label{system}
%\end{equation}

Numerical solutions of Eqs.~\eqref{system} and \eqref{boundary_cond} can be easily found and 
the corresponding layer displacements $\zeta_1(x)$ are represented in Fig.~\ref{fig:3Dscheme} by solid lines. It is instructive to obtain an approximate analytical solution of Eqs.~\eqref{system} and \eqref{boundary_cond}. To that end, we start from the Schwerin's classic solution valid at the 
specific Poisson's ratio $\sigma=1/3$~\cite{Komaragiri:2005wn},
\begin{equation}
    h_0(x) = \left(\frac{9x}{16}\right)^{1/3},
\end{equation}
and seek the solution in the form of $h(x) = h_0(x)+ px^{t}$, where $p$ and $t$ are the parameters. Substituting such expression into Eqs.~\eqref{system} and \eqref{boundary_cond} we arrive at
\begin{equation}
    h(x) = \sqrt[3]{\frac{9}{16}}\left(x^{1/3} - c^2(\sigma)x^{5/3}\right) + \mathcal O(x^3),
    \label{f_approx}
\end{equation}
with the function $c(\sigma)$ given by
\begin{equation}
    c^{2}(\sigma) = \frac{1/3 - \sigma}{5/3 - \sigma}.
\end{equation}
%The dots in Eq.~\eqref{f_approx} denote the terms with higher powers of $x$. 
The leading omitted term is $\propto x^{3}$ with a numerically small coefficient.
After substitution of the analytical expression from 
Eq.~(\ref{f_approx}) to the Eq.~(\ref{fg}) we obtain the shape of the layer
\begin{equation}
    \zeta_{1}(0) - \zeta_{1}(x) = \left(\frac{3a^2P}{\pi Y}\right)^{1/3}\frac{\tanh^{-1}\left[c(\sigma)x^{2/3}\right]}{c(\sigma)}.
    \label{form}
\end{equation}
The analytical expressions are plotted in Fig.~\ref{fig:3Dscheme} by dashed lines. 
One can see a very good agreement between the numerics and analytical approximations. 
Note that an interplay of the \rev{temperature-induced} flexural fluctuations 
of the layers and anharmonicity can result in renormalization of $B$ and $Y$ 
\rev{for sufficiently small wavevectors, $q<q^{*}$ with the temperature-dependent $q^{*}$}, 
see appendix~\ref{app:fluct:renorm} for details.

According to Ref.~\cite{Komaragiri:2005wn}, the deflection of the layers \rev{satisfies a self-consistency
condition:}
\begin{equation}
\label{nonlin:def}
    \zeta_{1}(0) = -\zeta_{2}(0) = 
    \frac{\zeta(0)}{2} = \mathcal F(\sigma)\left(\frac{a^2P}{Y}\right)^{1/3},
\end{equation}
\rev{where $\zeta(0)\equiv\,\zeta_1(0)-\zeta_2(0)$ is the relative displacement of the layers.  Note that, $P$ depends on $\zeta(0)$ according to Eq.~\eqref{force} making Eq.~\eqref{nonlin:def} strongly nonlinear.} % equivalent to a strongly nonlinear fifth-power in $\zeta(0)$ algebraic equation.}

The function $\mathcal F(\sigma)$ can be found numerically. Our analytical approximation 
follows from Eq.~\eqref{form} at $x=1$ as $\mathcal F(\sigma) = {\sqrt[3]{3/\pi}}\tanh^{-1}[c(\sigma)]/c(\sigma).$
Reference~\cite{Komaragiri:2005wn} presents the fit to numerical result in the form $\mathcal F(\sigma) = 1.0491 - 0.1462\sigma - 0.15827\sigma^2$. Our analysis shows that the difference between numerical calculation and analytical one does not exceed $1\%$ in the range of Poisson's ratio variation from $\sigma=0$ to $\sigma=1/2$. Also, the function $\mathcal F(\sigma)$ remains quite close to unity in this range varying from $\mathcal F(0) \approx 1.05$ to $\mathcal F(1/2) \approx 0.94$.

% \commentMisha{as far as I remember ``our'' asymptotics is slightly different. We can present our asymptotics and their numerical fit and say that they are also quite close.} \commentZakhar{``Our'' asymptotics is Eq.~\ref{form} with $x=1$. It is also good enough, however Komarigi's is slightly better Fig.~\ref{w(nu)}.} 
% \begin{figure}[ht]
%     \centering
%     \includegraphics[width=\linewidth]{image.png}
%     \caption{Deflection 
%     \commentMisha{Do we need it?}}
%     \label{w(nu)}
% \end{figure}

\begin{figure}[ht]
    \centering
    \includegraphics[width=\linewidth]{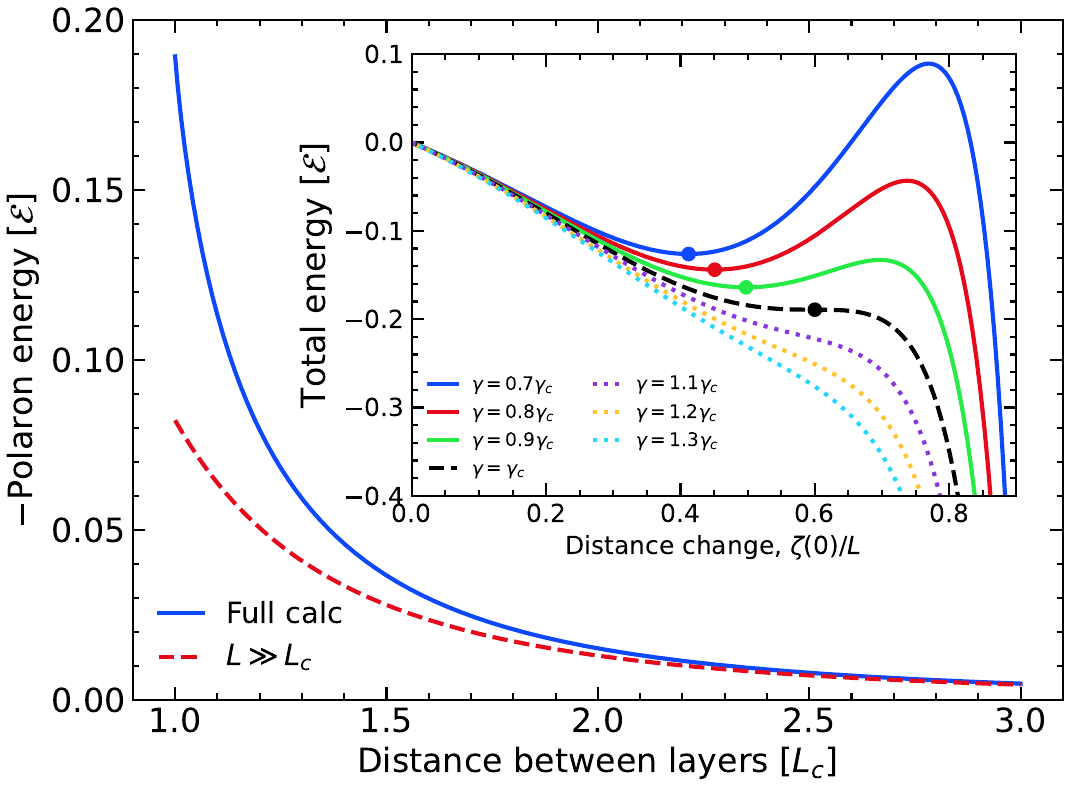}
    \caption{Absolute value of the exciton-polaron binding energy in the anharmonic regime as function of the distance between layers in the units of critical distance $L_c$, Eq.~\eqref{Lc}. Solid blue curve has been calculated after minimization of Eq.~(\ref{energy_l}), dashed red curve represents its asymptotic behavior at large distance Eq.~(\ref{E_nl_asympt}). Inset demonstrates the total energy of the system as function of distance between the layers change for different interaction parameters $\gamma$ in Eq.~(\ref{energy_l}). Solid curves demonstrate energy for the realizations, in which polaron exists. The polaron energies shown by dots are obtained by minimization of Eq.~(\ref{energy_l}). Dashed black line shows the critical behavior with maximum interaction of polaron existence. Dotted curves represent large interaction with monotonic behavior of the energy, when the layers stick and the polaron collapses. }
    \label{fig:E_nl}
\end{figure}

The elastic energy of the layer under the external force $P$ is the work of this force in the process of quasi-static stretching of the film
\begin{equation}
    {U}(\zeta_1{(0)}) = \int_0^{\zeta_1{(0)}}P(\zeta_1)d\zeta_1 = \frac{Y\zeta_{1}^{4}(0)}{4\mathcal F^{3}(\sigma)a^2}.
    \label{elastic}
\end{equation}
Making use of the Coulomb law, Eq.~\eqref{force}, we obtain the total energy of the system, i.e., the energy of the exciton-polaron with the anharmonic deformations:
\begin{equation}
    E = \frac{e^2}{L} - \frac{e^2}{L -\zeta(0)} + 2U\left(\frac{\zeta(0)}{2}\right).
    \label{energy_general}
\end{equation}
 Combining Eqs.~\eqref{elastic} and \eqref{energy_general} we get the total energy as function of center displacement $\zeta(0)$
\begin{equation}
    E = %\frac{e^2}{L}\left(\frac{\zeta(0)}{L+\zeta(0)} + \frac1{32\tau}\frac{\zeta(0)^4}{L^4}\right) \\ =
    \mathcal E\gamma^{1/5}\left(-\frac{\zeta(0)}{L-\zeta(0)} + \frac1{32\gamma}\frac{\zeta{^4}(0)}{L^4}\right),
    \label{energy_l}
\end{equation}
where 
\begin{equation}
\label{energ:const}
\mathcal E = \left(\frac{e^8Y}{{\mathcal F}^{3}(\sigma)a^2}\right)^{1/5},
\end{equation}
and we introduced the interaction parameter $\gamma$ as
\begin{equation}
    \gamma = \frac{{\mathcal F}^{3}(\sigma)a^2e^2}{YL^5}.
\end{equation}

The energy $E$ in Eq.~\eqref{energy_l} for different values of interaction parameter $\gamma$ is shown in inset in Fig.~\ref{fig:E_nl}. For relative displacement $\zeta(0)\to L$ the Coulomb energy dominates and $E\to -\infty$. Depending on the value of $\gamma$ an additional minimum in the $E[\zeta(0)]$ appears, as shown by dots in the inset. The exciton-polaron energy corresponds to the local minimum, which can be found as a root of derivative $\partial E / \partial \zeta(0)=0$, and plotted in the main panel of Fig.~\ref{fig:E_nl} by the solid line. The bound state exists only for sufficiently small interaction parameter $\gamma$ being less than critical parameter $\gamma_c$ 
determined by the conditions $\partial E / \partial \zeta(0)=0,$ $ \partial^{2}E/\partial\zeta^{2}(0)=0$ for the
same $\zeta(0),$ producing the inflection point in the $E[\zeta(0)]$-dependence corresponding to the dot 
at the dashed black line in the inset of Fig. \ref{fig:strong_coupling}. Using Eq.~\eqref{energy_l}, we obtain: 
\begin{equation}
    \gamma < \gamma_c = \frac{3^3\cdot 2^4}{10^5} = 0.00432, \qquad \zeta(0) < 0.6L,
\end{equation}
which means that layers are rigid enough. For stronger interaction, the local minimum 
is absent and the layers stick together due to the Coulomb interaction.

% \begin{equation}
%     w(0) < 0.3L, \qquad \frac{3^3\cdot 2^4}{10^5f(\nu)^3}\frac{YL^5}{a^2e^2} = \frac{0.00432}{f(\nu)^3}\frac{YL^5}{a^2e^2} > 1
% \end{equation}

For sufficiently weak interaction, $\gamma\ll \gamma_c$, the dependence of the Coulomb force on displacement can be neglected. Setting $P = e^2/L^2$ we obtain the following expression for the polaron energy
\begin{equation}
    E = -\frac3{2}\mathcal F (\sigma)\left(\frac{a^2e^8}{YL^8}\right)^{1/3} = {-\frac{3}{2}\mathcal{E}}\gamma^{8/15}.
    \label{E_nl_asympt}
\end{equation}

\rev{Note that if the exciton Bohr radius $a_{B}$ is 
larger as compared to the interlayer distances, but still much smaller than the flake size ($L \ll a_{B} \ll a$), 
than $P = D = 4e^2/a_{B}^2$ in accordance with Eq.~\eqref{D:est} and instead of Eq.~\eqref{E_nl_asympt} 
we have $E =-6\mathcal{F}(\sigma)\left[(4a^2e^8)/(Ya^{8}_{B})\right]^{1/3}$. 
Divergence at $\zeta(0) = L$ disappears and the polaron 
energy as the function of $\zeta(0)$ corresponding to the inset of Fig.~\ref{fig:E_nl} becomes a function with a single extremum.}

Exciton-polaron binding energy $E$ together with its asymptotic behavior Eq.~(\ref{E_nl_asympt}) as function of distance between layers $L$ is shown in Fig.~\ref{fig:E_nl}. Here $L$ is taken in units of critical distance $L_c$, which correspond to distance at $\gamma = \gamma_c$:
\begin{equation}
    \label{Lc}
    L_c = \left(\frac{{\mathcal F}^{3}(\sigma)a^2e^2}{Y\gamma_c}\right)^{1/5}.
\end{equation}

% \begin{figure}[ht]
%     \centering
%     \includegraphics[width=\linewidth]{w(L).pdf}
%     \caption{w(L)}
%     \label{fig:E(L)}
% \end{figure}

% \begin{figure}[ht]
%     \centering
%     \includegraphics[width=\linewidth]{w(L).pdf}
%     \caption{w(L)}
%     \label{fig:w(L)}
% \end{figure}

\section{Discussion}\label{sec:discuss}

Here we discuss the obtained results and highlight the key aspects of the interlayer exciton-polaron.

\subsection{Strong strain and lattice distortion}

%Interlayer exciton-polarons in bilayer systems of atomically thin semiconductors such 
%as the transition metal dichalcogenides are formed by the exciton coupling to the 
%out-of-plane displacements of the layers. Here we considered two important aspects of the physics of %interlayer exciton-polarons.

In this subsection we consider the ability to manipulate the properties of the 
exciton-polaron by applying a strain to 
the layers. As it has been shown~\cite{https://doi.org/10.1002/andp.202000339}, 
in the limit of zero cut-off frequency ($\omega_0 \to 0$), the polaron collapses due to softness
of the flexural phonons modes. The strain naturally increases the stiffness of these modes and, thus, works against the
collapse, however, without cut-off frequency layers still stick. We studied this effect in Sec.~\ref{sec:linear} and obtained modification of the polaron energy and  effective mass due to the strain effect. 

To understand realization of the strain-induced effects, we estimate the tension $T$ in terms of the material parameters as: 
\begin{equation}\label{T_Y_a0}
T\sim Y\frac{\delta a_{0}}{a_{0}},  
\end{equation}
where the Young's modulus $Y\sim\,W/a_{0}^{2},$ where 
$W\sim e^{2}/a_{0} \sim 1$~eV is the characteristic electron energy bandwidth, 
$a_{0}$ is the lattice constant, and $\delta a_{0}$ is its variation due to applied external force. Next, we consider the polaron
formed by an exciton. Aiming at analysis of the limiting cases, we introduce parameter $l_{\max}\equiv\max(a_{B},L)$, maximal value between exciton Bohr radius, $a_B$, and the interlayer distance, $L$, which roughly determines the strength of the exciton coupling with out-of-plane vibrations $D\sim e^{2}/l_{\max}^2$, cf. Eq.~\eqref{D:est}. Thus, for sufficiently large tension where the interlayer coupling can be disregarded, the equilibrium condition has the form: 
\begin{equation}\label{rel}
\zeta \sim \frac{D}{T}. 
\end{equation}%
The relation \eqref{rel} describes the balance of the exciton-induced pressure and tension, see Eq.~\eqref{general}. 
Accordingly, the polaron energy can be estimated as 
\begin{equation}
    \delta E \sim - D \zeta \sim -\frac{D^{2}}{T},
\end{equation}
which is in agreement with the second line of Eq.~\eqref{asympt} if one omits logarithmic factor unimportant for the simple estimates.

We begin with the possibility of realization of large-strain condition, $\alpha>1.$ Following the definition in Eq. \eqref{beta_alpha_K} and Eq. \eqref{T_Y_a0}, we obtain relative lattice distortion corresponding to 
$\alpha>1$ as:
\begin{equation}
\frac{\delta a_{0}}{a_{0}} > \frac{M}{m}\frac{\hbar \omega _{0}}{W}.
%\frac{\hbar \omega _{0}}{e^{2}/l_{\max}}
\end{equation}%
Note that $\alpha>1$ condition can also be recast as $ms^2/(\hbar\omega_0) >1$ with $s = \sqrt{T/\rho}$ is the tension-induced sound velocity. Thus, one needs to minimize $\hbar \omega _{0}$ to achieve large $\alpha.$ 
Taking into account that $M/m\sim\,10^{5},$ $\omega
_{0}$ behaves as $1/L^{3},$ and at $L\sim a_{0}$,  $\hbar\omega_{0}\sim 3\mbox{ meV},$~\cite{Jeong:2016aa}, we obtain that one needs $L\sim\,10 a_{0}$ to produce an achievable strain to overcome the effect of the van-der-Waals forces. 

Next, we establish the boundary of the strong tension regime formulated in Eq. \eqref{strong:zeta_lim} as $v\gg\sqrt{\beta}$ as applied in Eq. \eqref{strong:energy}. Taking into account the definition of the system parameters in Eqs. \eqref{varkappa_nu} and \eqref{beta_alpha_K}, we estimate $v$ as $\sim T/(\omega_{0}\sqrt{B\rho}).$ Since typical values of $B$ 
are of the order of $W,$ we obtain $v\sim\sqrt{W/\rho\omega_{0}^2}$.  As a result, strong tension effects begin at relative distortions $\delta\,a_{0}/a_{0}\sim (|E_{B}|/W)\times{(a_{0}/l_{\max})}.$  Taking into account that at $a_{0}/l_{\max}\sim 0.1$ one has $|E_{B}|/W\sim 0.1$ we obtain the resulting $\delta\,a_{0}/a_{0}\sim 10^{-2}$ and the tension $T\sim 10\mbox{ erg/cm}^{2}.$ 
Note that condition $v>\sqrt{\beta}$ requires a much smaller lattice distortion than the condition
$\alpha>1,$ implying that a large effect of the tension can be achieved at a relatively small $\alpha,$
as can be seen from the analysis in the Appendix B.

\subsection{Critical size of the flake and polaron collapse}

There is another aspect, which is related to the nonlinear regime in the polaron formation. The layer deformations  become so large that the coupling between the out-of-plane and in-plane displacements starts to play a role. The effect of the nonlinearity increases with the 
increase of the flake size $a$, and the polarons collapse due to sticking of the layers 
if the flake size is sufficiently large. 
First, we mention that by comparing linear,  $\sim\,B\zeta^{2}/a^{2},$
and nonlinear, $\sim\,W\zeta^{4}/(a_{0}^{2}a^{2})$ contributions to the deformation energy 
and taking into account that $B\sim\,W$ we see that the nonlinear term becomes important at deformations $\zeta>a_{0}.$

We assume here the conditions considered in Sec. \ref{sec:anh} with $l_{\max}=L.$ 
For nonlinear deformation one has from Eq.~\eqref{nonlin:def}: 
\begin{equation}
\zeta \sim \left( \frac{e^{2}}{L^{2}}\frac{a^{2}}{Y}\right)^{1/3}. 
\end{equation}
Thus, taking into account that $Y\sim{W}/{a_{0}^{2}}$ with $W\sim\,e^{2}/a_{0},$
the deformation can be expressed as: 
\begin{equation}
\zeta 
%\sim\left( \frac{e^{2}}{l_{\max}^{2}}\frac{a^{2}}{e^{2}/a_{0}}a_{0}^{2}\right)^{1/3}
\sim a_{0}\left( \frac{a^{2}}{L^{2}}\right) ^{1/3}, \label{zeta:55}
\end{equation}
and depends solely on the dimensions of the system meaning that the nonlinear effects
become relevant at $a>L.$ The condition of collapse $\zeta \sim L$ implies 
\begin{equation}
\frac{a}{a_{0}}\sim \left(\frac{L}{a_{0}}\right) ^{5/2},
\end{equation}
suggesting that  the critical radius of the flake is rather small, being $\sim 100 a_{0}$
at $L\sim 5 a_{0}.$ 

\rev{For the $a_{B}\gg\,L$ realization one obtains $\zeta\sim\,a_{0}\left(a^{2}/a_{B}^{2}\right)^{1/3}$ 
and, correspondingly condition for sticking of the layers as $a\sim\,\left(L/a_{0}\right)^{3/2}a_{B},$ of the order
of $10a_{B}$ at $L\sim 5 a_{0}.$}

Interestingly, the physics described above can be realized in wide spectrum of the 
systems involving atomically-thin semiconductors and membranes. For example, one can 
consider a monolayer suspended over a dielectric or metallic substrate. 
In this situation, attraction of a charged donor to its electrostatic image can result 
in the monolayer deformation and donor-polaron effect. Depending on the system parameters, the gain in the polaron energy can, in principle, exceed the binding energy of electron to the donor rendering it in the autoionized state. The polaron effects may also affect the shape of the bubbles~\cite{Khestanova:2016wr} arising in two-dimensional crystals deposited on substrates. A membrane with embedded ions, which, due to the interaction with image charges, produce the membrane deformation, could be another example of the polaron-related effects.

\section{Conclusion}\label{sec:concl}

We have theoretically studied exciton-polarons in bilayers of transition metal dichalcogenides formed by excitons coupled to the flexural phonon modes. 
We demonstrated the ability to manipulate the exciton-polaron energies and effective masses by applying strain to the layers in small and large strain regimes and also for weak and strong 
exciton-phonon coupling. Application of the strain decreases electron-phonon coupling and diminishes its effect on the 
polaron energy and effective mass. It is due to hardening of the flexural phonons in the presence  
of the strain. The role of strain increases with increasing the distance between the layers since the out-of-plane phonon frequency decreases
rapidly with the interlayer distance.

We also studied nonlinear regime of the polaron formation and obtained the limits of applicability 
of this approach. As expected, the nonlinear effects become important if the lattice flexural displacement in the polaron
exceeds the lattice constant. The effect  of the nonlinear coupling strongly depends on the size of the nanoflake
where the exciton is located and at a sufficiently large flake the exciton collapses and layers stick to each other. 
These results can be useful for the design of transport and optical properties of bilayers and their nanostructures.   

\acknowledgements

Z.A.I. acknowledges support of the Russian Science Foundation project \#20-42-04405 for numerical calculations 
and the Foundation for the Advancement of Theoretical Physics and Mathematics ``BASIS''.
E.Y.S. acknowledges support of the Spanish Ministry of Science and the European Regional Development
Fund through PGC2018-101355-B-I00 (MCIU/AEI/FEDER, UE) Grant and the Basque Country Government 
through Grant No. IT986-16.

\appendix

\section{Effects of finite temperature, fluctuations and anharmonicity}\label{app:fluct:renorm}

\rev{To discuss the role of a nonzero temperature, we begin by noticing that with the increase in the temperature 
both virtual and real phonon-assisted processes  
take place and can become essential for the understanding of the effects considered 
in terms of the perturbation theory. The analysis performed in Ref.~\cite{https://doi.org/10.1002/andp.202000339} shows that at 
temperatures $k_{B}T \gtrsim \hbar\omega_{0}$ (where $k_{B}T$ is the temperature in the energy units) the 
constant $\beta$ in Eq.~\eqref{pol:energ:weak} increases approximately 
by a factor of $2k_B T/\hbar\omega_0$. Also, exciton-phonon scattering starts to play a role 
resulting in temperature-induced broadening of the polaron state.}

\rev{Another effect of finite temperature is its impact on the flexural 
phonon dispersion at small wavevectors, caused by the anharmonicity in the motion of the layers.} 
For a single layer the correlation function of the out-of-plane displacements severely 
diverges in the limit of $q\to 0$ with $\langle \zeta_q^2\rangle \propto k_B T/q^4$. 
The coupling of the flexural and in-plane vibrations results in the wave-vector 
dependent renormalization of the bending rigidity yielding steeper dispersion 
of the flexural phonons 
in the form~\cite{nelson:2004aa,PhysRevB.92.155428,Gornyi:2016aa,PhysRevB.94.195430,Le-Doussal:2018vk,Katsnelson2020}
\begin{equation}
    \label{disper:renorm}
    \omega_q \propto q^{2-\eta/2}, \quad q\lesssim q^{*},
\end{equation}
where $\eta \approx 0.6\ldots 0.8$ is the exponent describing the renormalization 
of the bending rigidity, and $q^{*}\sim \sqrt{k_B T Y/B^2}$ serves as an inverse 
critical length scale of the theory of phase transitions. At $q\gtrsim q^{*}$ the quadratic 
dispersion of flexural phonons is restored. Thus, if the system size is sufficiently 
large, $aq^{*}\gg 1$, the effects of the phonon dispersion renormalization 
described by Eq.~\eqref{disper:renorm} should be taken into account. 
Still, the phonon dispersion remains sufficiently soft and, in the absence 
of the cut-off frequency $\omega_0$, the polaron energy shift diverges already 
in the weak-coupling regime, cf. Eq.~\eqref{2nd:total}. The results for the 
polaron shift and effective mass obtained with account for finite $\omega_0$ are similar to those presented in the main text.

The coupling between the out-of-plane and in-plane vibrations 
also result in the anomalous elasticity of the layers~\cite{Los:2017tz,PhysRevE.101.033005}. 
For $aq^{*}\gg 1$ the analysis of the anharmonic regime presented in Sec.~\ref{sec:anh} 
remains valid, but both bending rigidity and Young's modulus should be replaced by 
the renormalized values~\cite{Los:2017tz,PhysRevE.101.033005}
\begin{equation}
    \label{BY:renorm}
 B\to B (aq^{*})^\eta, \quad Y\to Y(aq^{*})^{-\eta_u},   
\end{equation}
with $\eta_u \approx 2-2\eta$.

We note that the anharmonic coupling between the flexural and in-plane 
phonons can be strongly suppressed if the interaction between the layers is large enough. 
Indeed, the frequency $\omega_0\ne 0$ suppresses the divergence 
of $\langle \zeta_q^2\rangle$ at $q\to 0$ validating the analysis presented in the main text.
The strong role of this suppression can be seen from the following estimate. Following 
Eq. \eqref{disper}, one obtains $\varkappa (q^{*})^{2}/\omega_{0}\sim \sqrt{m/M}\times(k_{B}T)/(\hbar\omega_{0})\ll 1$ due to a small $m/M$ ratio. Therefore, the temperature-dependent phonon dispersion renormalization cannot modify the effect of finite $\omega_{0}.$

\section{Layers shape in the strong coupling regime}

\begin{figure}[hb]
    \centering
    \includegraphics[width=\linewidth]{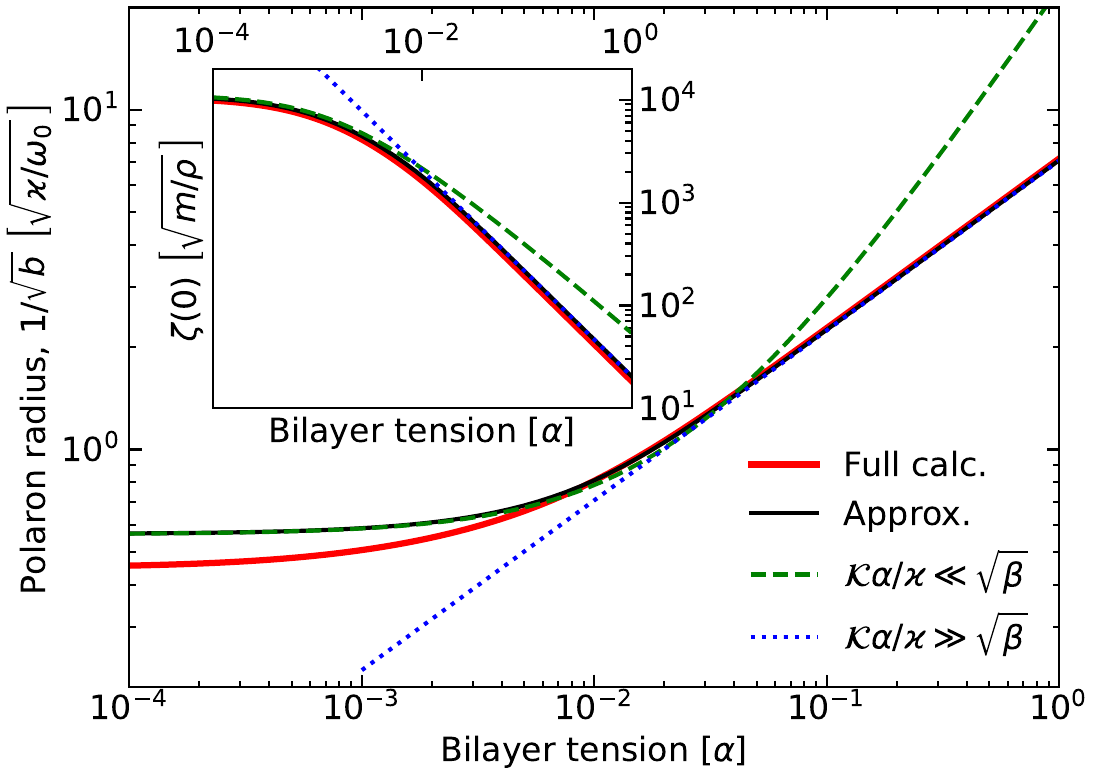}
    \caption{Polaron radius in the strong coupling regime in the units $\sqrt{\varkappa / \omega_0}$ as a function of the bilayer tension. Solid black curve has been calculated after minimization Eq.~\eqref{strong:exact}. Solid red curve is found after substitution of $b$ from Eq.~\eqref{strong:b} in Eq.~\eqref{strong:r}. Dashed green and dotted blue curves show, respectively, small and large tension asymptotics Eq.~\eqref{strong:r}. Inset demonstrates the deflection of the polaron in the units of $\sqrt{m / \rho}$ as the function of bilayer tension. Solid black curve has been calculated after minimization Eq.~\eqref{strong:exact}. Solid red curve is found after substitution of $b$ from Eq.~\eqref{strong:b} in Eq.~\eqref{strong:zeta(0)}. Dashed green and dotted blue curves show, respectively, small and large tension asymptotics Eq.~\eqref{strong:zeta_lim}.}
    \label{fig:strong_sizes}
\end{figure}

In the linear regime layers can be described as infinite planes, while the polaron has a finite size. 
Naturally, the parameter $b$ in Gaussian trial function in Eq.~\eqref{trial:gauss} is related to the in-plane size of the polaron $R$ as $R=1/\sqrt{b}$. After minimization of the energy in Eqs.~\eqref{strong:exact} and~\eqref{strong:E(b)} it can be written as:
\begin{align}
    b 
    &\approx \left(\sqrt{\frac{\beta}{2\pi} - u} - v\right)\frac{\w_0}{4\varkappa} \nonumber \\
    &\approx \dfrac{\w_0}{4\varkappa}\begin{cases}
        \sqrt{\dfrac{\beta}{2\pi}} - v, & v \ll \sqrt{\beta} \\
        \\
        \dfrac{\beta}{4\pi v}, & v \gg \sqrt{\beta}
    \end{cases}.
    \label{strong:b}
\end{align}
Then, a good estimate of the polaron radius is
\begin{equation}
    R  \approx \begin{cases}
    2\sqrt{\dfrac{\varkappa}{\omega_0}}\left[\left(\dfrac{2\pi}{\beta}\right)^{1/4} + \left(\dfrac{2\pi}{\beta}\right)^{3/4}\dfrac{v}{2}\right], & u \ll \sqrt{\beta}, \\
    \\
        4\sqrt{\dfrac{\pi v\varkappa}{\beta\omega_0}}, & u \gg \sqrt{\beta},
    \end{cases}
    \label{strong:r}
\end{equation}
and shown in Fig.~\ref{fig:strong_sizes}. In the linear regime the out-of-plane displacement of the layers is proportional to the force, $\zeta_{\bm q} = 2f_{\bm q} / (\rho\omega_q^2)$, and can be found as:
\begin{equation}
    \zeta(r) = \sum_{\bm q}\zeta_{\bm q}e^{i\mathbf{qr}} = \frac{D}{\pi\rho}\int_0^\infty\frac{\exp[-q^2 / (8b)]J_0(qr)qdq}{\omega_0^2 + \nu q^2 + \varkappa^2q^4}.
    \label{strong:deflection}
\end{equation}
Note that the polaron radius $R$ is significantly smaller, than the size of the deformed area. The latter is connected with the scale where $\zeta(r)$ in Eq.~\eqref{strong:deflection} significantly decreases. In the absence of tension deformation radius can be estimated as $R_d \sim \sqrt{\varkappa} / \omega_0$ according to $R_d \sim 1/q_d$ where the denominator in Eq.~\eqref{strong:deflection} starts to increase. This result can be easily understood assuming that the cut-off frequency $\omega_0$ is related to the in-plane size of the layer $a$. In this situation, under a point load, the layer deformation extends over the whole layer~\cite{ll7_eng}.

\begin{figure}[ht]
    \centering
    \includegraphics[width=\linewidth]{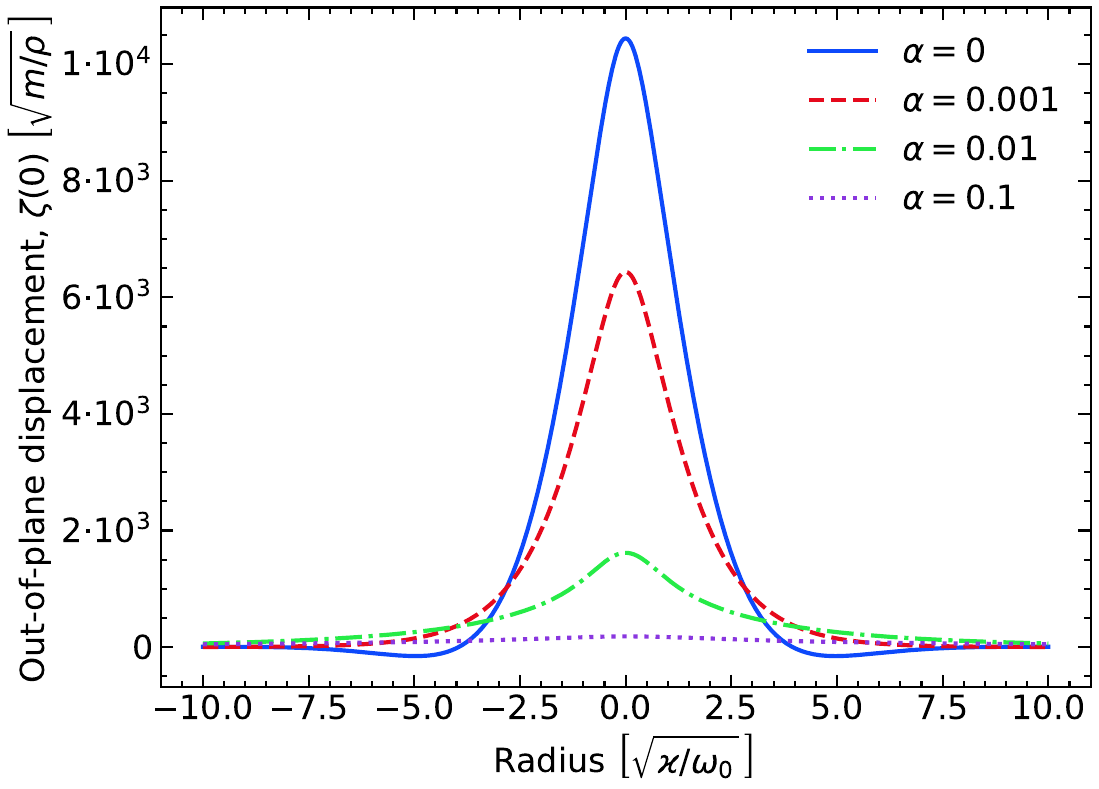}
    \caption{Shape of the layers calculated after Eq.~\eqref{strong:deflection} for different bilayer tension $\alpha$ (presented in the plot legend).}
    \label{fig:strong_shape}
\end{figure}

With the increase in the tension deformation size is determined by tension, $R_d \sim \omega_0 / \sqrt{\nu}$, the layers become stiffer and the polaron effect diminishes: the polaron size increases, while the layer deflection decreases. The shape of the layers found from Eq.~\eqref{strong:deflection} is presented in Fig.~\ref{fig:strong_shape}.

After transformation similar to one between Eq.~\eqref{strong:exact} and Eq.~\eqref{strong:E(b)}, the layer deflection at the polaron center is expressed as
\begin{multline}
    \zeta(0) \approx  \frac1{2\pi}\frac{\mathcal{K}}{\varkappa}\sqrt{\frac{2\beta m}{\rho u}} \\ \times\left(\arctan\frac{\sqrt{u}}{v} - \arctan\frac{\sqrt{u}}{v+8\varkappa b / \omega_0}\right).
    \label{strong:zeta(0)}
\end{multline}
The limits of Eq.~\eqref{strong:zeta(0)} for the small and large tension are
\begin{equation}
    \zeta(0) \approx \frac1{2\pi}\dfrac{\mathcal{K}}{\varkappa}\sqrt{\frac{2\beta m}{\rho}}\begin{cases}
        \dfrac1{\sqrt{u}}\arctan\dfrac{\sqrt{u}}{v}, & v \ll \sqrt{\beta}, \\
    \\
        \dfrac1{2v}\ln{\dfrac{\beta}{\pi}}, & v \gg \sqrt{\beta}.
    \end{cases}
    \label{strong:zeta_lim}
\end{equation}
The out-of-plane displacement of the center as function of bilayer tension is shown in inset in Fig.~\ref{fig:strong_sizes}.

The system can be correctly described in linear regime only if the angle of the layers deflection is sufficiently small. This criteria can be represented in terms of deformation radius $R_d$ and layer deflection $\zeta(0)$:
\begin{equation}
    \left|\frac{d\zeta}{dr}\right| \approx \frac{\zeta(0)}{{R_d}} \approx {\frac{\mathcal K}{4\varkappa}\sqrt{\frac{2\beta m\omega_0}{\rho\varkappa}}}.
    \label{zeta_derivative}
\end{equation}
\vspace{0.1cm}

\noindent Equation~\eqref{zeta_derivative} is valid at $\alpha\to 0$. In the presence of the tension, the angle decreases. 

Let us analyze the dependence of the deflection angle on the interlayer distance at $L\gg a_B$. If $\omega_0$ is related to the size of the layers and is independent of the interlayer distance $L$ then $\beta\propto 1/L^4$ and the angle decreases with the increase in the interlayer distance as $|d\zeta / dr| \propto 1/L^2$. In the case of the van der Waals coupling between the layers,  we have $\beta \propto L^2$ and $\omega_0 \propto 1/L^3$, thus, the deflection angle decreases with increasing of the interlayer distance as $|d\zeta / dr| \propto 1/\sqrt{L}$.

The theory above is valid at $|d\zeta/dr| \ll 1$. This condition should be fulfilled both in the harmonic and anharmonic regimes, Sec.~\ref{sec:linear} and \ref{sec:anh}, respectively. Note that the transition between the linear and anharmonic regimes is determined, in agreement with Eqs.~\eqref{anharmonic:limit} and \eqref{zeta:55}, by $B^3 \sim a^4P^2Y$.

%\bibliography{all-1}
%merlin.mbs apsrev4-1.bst 2010-07-25 4.21a (PWD, AO, DPC) hacked
%Control: key (0)
%Control: author (0) dotless jnrlst
%Control: editor formatted (1) identically to author
%Control: production of article title (0) allowed
%Control: page (1) range
%Control: year (0) verbatim
%Control: production of eprint (0) enabled
%

\end{document}